\documentclass[letterpaper,12pt]{JHEP3}
\usepackage{graphicx}
\usepackage{rotating}
\usepackage{latexsym}

\newcommand{\beq}{\begin{equation}}
\newcommand{\eeq}{\end{equation}}
\newcommand{\bqa}{\begin{eqnarray}}
\newcommand{\eqa}{\end{eqnarray}}
\newcommand{\npd}{\nabla_\perp}
\newcommand{\npu}{\nabla^\perp}
\newcommand{\sn}{s}
\newcommand{\mt}{\left(\nabla\cdot u\right)}

\author{Paul Romatschke\\
Institute for Nuclear Theory, University of Washington, \\
Box 351550, Seattle, WA, 98195\\
E-mail: \email{paulrom@phys.washington.edu}}

\title{Relativistic Viscous Fluid Dynamics and Non-Equilibrium Entropy}

\abstract{
Fluid dynamics corresponds to the dynamics of a substance in the long wavelength limit.
Writing down all terms in a gradient (long wavelength) expansion
up to second order for a relativistic system at vanishing charge density, one
obtains the most general (causal) equations of motion for a fluid in the presence
of shear and bulk viscosity, as well as the structure of the non-equilibrium entropy 
current.
Requiring positivity of the divergence of the non-equilibrium entropy current
relates some of its coefficients to those entering the equations of motion.
I comment on possible applications of these results for conformal and non-conformal
fluids.}

\preprint{INT-PUB-09-032}
\date{\today}
\begin{document}


\section{Introduction}

The theory of fluid dynamics has a long history, starting with its inception
by L. Euler in 1755 \cite{Euler} and generalized to viscous fluids by the works
of C. Navier \cite{Navier} and G. Stokes \cite{Stokes} in the 19th century. Besides being necessary
to realistically model the behavior of real fluids in many situations,
viscosity is known to be essential for the presence of laminar (or smooth)
flows since it dampens the turbulent instability generally inherent 
to ideal (non-viscous) fluids \cite{LL}\S26. As can be understood from the work
by Navier and Stokes, viscous effects correspond to (first order) gradients
of the equilibrium properties of the system (temperature, fluid velocity, etc.).

For relativistic systems, additional complications arise: a first
order gradient expansion \`a la Navier and Stokes leads to 
a set of fluid dynamics equations that allow faster-than-light
signal propagation, violating causality \cite{Kranys}. In the late 20th century,
motivated by questions in general relativity, 
M\"uller \cite{Mueller1}, Israel and Stewart \cite{IS0} showed that by including second
order gradient terms, the resulting fluid dynamics equations
could be made causal. In the last decade -- driven by 
nuclear physics experiments on relativistic heavy-ion collisions 
\cite{Adcox:2004mh,Back:2004je,Arsene:2004fa,Adams:2005dq} -- 
a program to determine the precise form of the equations 
for a relativistic viscous fluid was initiated 
\cite{Muronga:2003ta,Koide:2006ef,Baier:2007ix,Betz:2008me}. This program
has not been finished, and the present work is meant to 
provide another step towards its completion.

It should be noted that the question of causality is intimately
linked to the property of hyperbolicity of the fluid dynamic theory,
i.e. the question of whether a well-defined initial value formulation
can be given. While for small perturbations around equilibrium
hyperbolicity and causality can be shown for various second order theories,
it seems that in the case of strong non-linear out of equilibrium situations
hyperbolicity has been rigorously investigated only in so-called
divergence type theories \cite{Geroch:1990bw}, e.g. M\"uller's theory
\cite{Mueller1}. In particular, the causality properties of 
the theory by Israel and Stewart far out of equilibrium remain unknown.
This work is based on the more recent interpretation of 
fluid dynamics as an effective theory of the long-wavelength modes of the system,
by construction precluding any application to systems that are far from equilibrium.
While the effective theory specified below will share with 
other second theories the property
of causality (and hyperbolicity) for small perturbations around equilibrium, 
a proof of even hyperbolicity for strong perturbations
is beyond the scope of this work.

This article is organized as follows: in section \ref{sec:setup}, all 
relevant gradient structures up to second order 
are listed for a fluid at zero charge density,
and the concept of a conformal fluid is introduced. In section
\ref{sec:VEMT}, the most general form of the energy-momentum 
tensor for non-conformal fluids is presented, which fixes
the equations of motion for a relativistic viscous fluid.
Section \ref{sec:neentropy} deals with the most general form
of the entropy current for a relativistic fluid out of equilibrium
(again, at vanishing charge density), and its divergence.
In section \ref{sec:soknown}, I tried to collect all 
current knowledge about the coefficients multiplying second order 
gradient structures. Section \ref{sec:conclusions} contains
a discussion of the results and the conclusions.

\section{Setup}
\label{sec:setup}

Let us consider matter described by a relativistic quantum-field theory
at vanishing charge density (zero chemical potential). In equilibrium,
the long-wavelength dynamics of this system 
can be described by one scalar, one vector, and one tensor:
$$
\varepsilon, u^\mu, g^{\mu \nu}\,,
$$
which are the fundamental fluid dynamic variables
energy density $\varepsilon$, fluid four velocity $u^\mu$ and metric tensor
$g_{\mu \nu}$. The quantum field theory is supposed to furnish a
relation between the pressure $P$ and $\varepsilon$ via the equation of state,
$P=P(\varepsilon)$, giving rise to the speed of sound $c_s=\sqrt{dP/d\varepsilon}$
in the fluid.

Using these fundamental degrees of freedom, one can write
down the energy-momentum tensor of the fluid (see, e.g. \cite{LL}\S133),
\beq
T^{\mu \nu}_{\rm eq}=\varepsilon\,u^\mu u^\nu + P \Delta^{\mu \nu}\,,\quad \Delta^{\mu \nu}=g^{\mu \nu}+u^\mu u^\nu\,,
\eeq
where the subscript ``eq'' denotes equilibrium quantities containing
no gradients (zeroth order). Here and in the following the metric sign convention $(-,+,+,+)$ is used,
and all the calculations will be done in four space-time dimensions (though a generalization
to other number of dimensions should be straightforward). 
Without the presence of a source, the energy-momentum tensor is covariantly conserved, $\nabla_\mu T^{\mu \nu}=0$,
where $\nabla_\mu$ denotes the geometric covariant derivative. Projection of this equation
leads to the well-known equations of motion for an ideal relativistic fluid,
\bqa
u_\nu \nabla_\mu T^{\mu \nu}_{\rm eq}=- D \varepsilon- (\varepsilon+P) \nabla \cdot u 
&=& 0\,,\nonumber\\
\Delta_\nu^\alpha \nabla_\mu T^{\mu \nu}_{\rm eq}= (\varepsilon+P) D u^\alpha + \nabla_\perp^\alpha P &=&0\,,
\eqa
where the new notations 
$$D\equiv u^\mu \nabla_\mu\,,\quad \nabla_\perp^\mu \equiv \Delta^{\mu \nu} \nabla_\nu$$ 
for
the comoving time and space derivative were introduced. Using the basic thermodynamic
relations $\delta \varepsilon=T \delta \sn$, $\varepsilon+P=\sn T$ for the equilibrium entropy density $\sn$
and temperature $T$, the equations of motion for the ideal fluid become
\beq
D \ln \sn = - \nabla\cdot u\,,\qquad D u^\alpha = -c_s^2 \nabla_\perp^\alpha \ln \sn\,.
\label{eqds}
\eeq
This implies that not all first order gradients of the fundamental degrees
of freedom are independent: time derivatives may (up to higher order gradient corrections) always 
be recast as space derivatives. Therefore, to first order in gradients, the only independent
structures one can write down are $\npu_\alpha \ln \sn$ and  $\npu_\alpha u_\beta$ (no
coordinate-invariant first order gradient of the metric tensor exists). For later convenience,
gradients are sorted into three classes: scalars, vectors orthogonal to $u^\mu$, and symmetric traceless tensors
orthogonal to $u^\mu$. To first order in gradients, there is one of each class:
$$
(\nabla\cdot u)\,,\qquad \npu_\alpha \ln \sn\,, \qquad
\sigma_{\alpha \beta}\equiv\npu_{\alpha} u_\beta
+\npu_{\beta} u_\alpha-\frac{2}{3}\Delta_{\alpha \beta} (\nabla\cdot u)\,.
$$

To second order in gradients, the independent structures are 
$\npu_\alpha \npu_\beta \ln \sn$, $\npu_\alpha \npu_\beta u_\mu$, 
$\npu_\alpha \ln \sn \npu_\beta \ln \sn$,
$\npu_\alpha u_\mu \npu_\beta u_\nu$, $\npu_\beta u_\mu \npu_\alpha \ln \sn$ and the Riemann tensor (cf.~\cite{Wald}\S3.4)
$$
R^\lambda_{\ \mu \sigma \nu}\equiv\partial_\sigma \Gamma^\lambda_{\mu \nu}-\partial_\nu \Gamma^\lambda_{\mu \sigma}
+\Gamma^\kappa_{\mu \nu} \Gamma^\lambda_{\kappa \sigma}-\Gamma^\kappa_{\mu \sigma} \Gamma^\lambda_{\kappa \nu}\,,
$$
where 
$
\Gamma^\lambda_{\mu \nu}=\frac{1}{2} g^{\lambda \rho}\left(
\partial_\mu g_{\rho \nu}+\partial_\nu g_{\rho \mu} - \partial_{\rho} g_{\mu \nu}\right)\,.
$
Using the Ricci tensor $R_{\mu \nu}=R^\lambda_{\ \mu \lambda \nu}$ and Ricci scalar $R=R^\mu_\mu$,
one can build 7 independent scalars to second order in gradients,
\beq
\npu_\alpha \npd^\alpha \ln \sn\,,\quad
\npu_\alpha \ln \sn \npd^\alpha \ln \sn\,,\quad
\sigma_{\alpha \mu}\sigma^{\alpha \mu}\,,\quad
\Omega_{\alpha \mu}\Omega^{\alpha \mu}\,,\quad
\mt^2\,,\quad
u^\mu u^\nu R_{\mu \nu}\,,\quad
R\,,
\label{2scals}
\eeq
where $\Omega_{\mu \nu}=\frac{1}{2}\left(\npu_\mu u_\nu-\npu_\nu u_\mu\right)$ is the fluid 
vorticity which was used in the decomposition
$$
\npu_\mu u_\nu = \frac{1}{2}\sigma_{\mu \nu}+\Omega_{\mu \nu}+\frac{\mt}{3}\Delta_{\mu \nu}\,.
$$

Furthermore, one can build 6 independent vectors orthogonal to the fluid velocity:
\beq
\npu_\alpha \sigma^{\alpha \mu}\,,\quad
\npu_\alpha \Omega^{\alpha \mu}\,,\quad
\sigma^{\alpha \mu} \npu_\alpha \ln \sn\,,\quad
\mt \npd^\mu \ln \sn\,,\quad
\Delta^{\alpha \mu} u^\beta R_{\alpha \beta}\,,
\eeq
where one should note that $\Delta^{\alpha \mu} u^\beta R_{\alpha \beta}$ 
contains -- and therefore is used instead of -- the vector $\npd^\mu \mt$ (see appendix \ref{sec:appA}).

Finally, there are 8 independent symmetric traceless tensors orthogonal to the fluid
velocity. These have been found already in Ref.~\cite{Baier:2007ix} for the
case of conformal fluids (see below for the definition of a conformal fluid).
For a general relativistic fluid at vanishing charge density, the 8 independent
tensors are given by
\bqa
&\npd^{<\mu} \ln \sn \npd^{\nu>} \ln \sn\,,\quad
\npd^{<\mu} \npd^{\nu>}\ln\sn\,,\quad
\sigma^{\mu \nu} \mt\,,\quad
\sigma^{<\mu}_{\quad \lambda}\sigma^{\nu> \lambda}\,,&\nonumber\\
&\sigma^{<\mu}_{\quad \lambda}\Omega^{\nu> \lambda}\,,\quad
\Omega^{<\mu}_{\quad \lambda}\Omega^{\nu> \lambda}\,,\quad
u_\alpha u_\beta R^{\alpha <\mu \nu>\beta}\,,\quad
R^{<\mu \nu>}\,,&
\eqa
where for a second rank tensor $A^{\mu \nu}$
$$
\left<A^{\mu \nu}\right> = A^{<\mu \nu>}\equiv
\frac{1}{2}\Delta^{\mu \alpha} \Delta^{\nu \beta}\left(A_{\alpha \beta}+A_{\beta \alpha}\right)
-\frac{1}{3}\Delta^{\mu \nu}\Delta^{\alpha \beta}\,.
$$

\subsection{Conformal Fluids}

For certain situations it is advantageous to consider the
simplified case of a fluid without bulk viscosity. Since the bulk
viscosity coefficient $\zeta$ is related to the conformal anomaly $T^\mu_\mu$ via
a Kubo relation, a prime example for such a case is a system
that exhibits conformal symmetry, or covariance under local Weyl
rescalings of the metric:
\begin{equation}
g_{\mu \nu}\rightarrow e^{-2 w} g_{\mu \nu}\,,
\label{Weylres}
\end{equation}
where $w=w(x^\mu)$ is the local scale factor.
I will refer to the fluid description of a system that obeys
conformal covariance simply as ``conformal fluid''. An example for a
conformal quantum field theory is the ${\cal N}=4$ supersymmetric Yang-Mills theory
(SYM).
Note that in curved space the Weyl anomaly \cite{Duff:1993wm} would in general
break conformal symmetry, since, e.g., $T^\mu_\mu\propto R^2$.
However, this breaking occurs only at fourth order in gradients,
suggesting that for a fluid dynamic expansion up to third order
in gradients the Weyl anomaly can be ignored \cite{Baier:2007ix}.

Conformal covariance implies that objects transform
homogeneously under Weyl rescalings, meaning 
that gradients of the scale factor $w$ have to cancel.
To lowest order, consistency requires \cite{Baier:2007ix}
$$
\sn\rightarrow e^{3 w} \sn\,,\quad u^\mu \rightarrow e^w u^\mu\,.
$$
To first order in gradients, there are no scalars or vectors\footnote{For
non-vanishing charge density, parity-breaking effects may occur in the fluid. In this case,
also terms such as $\epsilon^{\mu \nu \alpha \beta}u_\nu \Omega_{\alpha \beta}$ are allowed.} 
but one symmetric tensor 
that transforms homogeneously under Weyl rescalings: $\sigma_{\mu \nu}$
(see appendix \ref{sec:weyl}). To second order in gradients, there are three conformal scalars,
\bqa
&{\cal S}_1=\sigma_{\mu \nu}\sigma^{\mu \nu}\,,\quad
{\cal S}_2=\Omega_{\mu \nu}\Omega^{\mu \nu}\,,&\nonumber\\
&{\cal S}_3=c_s^2\npu_\mu \npd^\mu \ln \sn+\frac{c_s^4}{2}\npu_\mu \ln \sn \npd^\mu \ln \sn
-\frac{1}{2}u_\alpha u_\beta R^{\alpha \beta}-\frac{1}{4}R+\frac{1}{6} \mt^2\,,&
\label{confscals}
\eqa
two conformal vectors orthogonal to $u^\mu$,
\beq
{\cal V}_1^\mu = \npu_\alpha \sigma^{\alpha \mu}+2 c_s^2 \sigma^{\alpha \mu}\npu_\alpha \ln \sn
-\frac{u^\mu}{2} \sigma_{\alpha \beta}\sigma^{\alpha \beta}\,,\quad
{\cal V}_2^\mu = \npu_\alpha \Omega^{\mu \alpha}+u^\mu \Omega_{\alpha \beta} \Omega^{\alpha \beta}\,,
\label{confvec}
\eeq
and five conformal symmetric traceless tensors orthogonal to $u^\mu$\,,
\bqa
{\cal O}_1^{\mu \nu}&=&R^{<\mu \nu>}-c_s^2 \left(2 \npd^{<\mu}\npd^{\nu>} \ln \sn
+\sigma^{\mu \nu}\mt-2 c_s^2
\npd^{<\mu} \ln \sn \npd^{\nu>} \ln \sn\right)\,,\nonumber\\
{\cal O}_2^{\mu \nu}&=&R^{<\mu \nu>}-2 u_\alpha u_\beta R^{\alpha <\mu \nu> \beta}\,,\nonumber\\
{\cal O}_3^{\mu \nu}&=&\sigma^{<\mu}_{\quad \lambda}\sigma^{\nu>\lambda}\,,\quad
{\cal O}_4^{\mu \nu}=\sigma^{<\mu}_{\quad \lambda}\Omega^{\nu> \lambda}\,,\quad
{\cal O}_5^{\mu \nu}=\Omega^{<\mu}_{\quad \lambda} \Omega^{\nu> \lambda}\,.
\label{conftens}
\eqa
These will be the building blocks of the energy-momentum tensor and entropy current
for conformal fluids.

\subsection{Non-Conformal Fluids}

For more general fluids that do not obey conformal invariance, all possible
gradients can contribute. In particular, 
to first order in gradients there are one scalar and vector,
\beq
\mt\,,\quad 
\nabla^\mu \ln \sn\,,
\eeq 
in addition to the tensor already found for the conformal case.

At second order, there are four additional scalars,
\beq
{\cal S}_4=\mt^2\,,\quad
{\cal S}_5=R\,,\quad
{\cal S}_6=\npu_\mu \ln \sn \npd^\mu \ln \sn\,,\quad
{\cal S}_7=u^\alpha u^\beta R_{\alpha \beta}\,,
\label{ncscals}
\eeq
four additional vectors,
\beq
{\cal V}_3^\mu=\Delta^{\mu \beta} u^\alpha R_{\alpha \beta}\,,\quad
{\cal V}_4^\mu=\sigma^{\mu \alpha} \npu_\alpha \ln \sn\,,\quad
{\cal V}_5^\mu=\Omega^{\mu \alpha} \npu_\alpha \ln \sn\,,\quad
{\cal V}_6^\mu=\mt \npd^\mu \ln \sn\,,
\eeq
and three additional tensors,
\beq
{\cal O}_6^{\mu \nu}=u_\alpha u_\beta R^{\alpha <\mu \nu>\beta}\,,\quad
{\cal O}_7^{\mu \nu}=\frac{\mt}{3}\sigma^{\mu \nu}\,,\quad
{\cal O}_8^{\mu \nu}=\npd^{<\mu}\ln \sn \npd^{\nu>}\ln \sn\,.
\label{nctens}
\eeq

\section{Viscous Fluid Dynamics: Energy-Momentum Tensor}
\label{sec:VEMT}

The energy-momentum tensor for a relativistic viscous fluid can be written as
$$
T^{\mu \nu}_{\rm non-eq}=T_{\rm eq}^{\mu \nu}+\Pi^{\mu \nu}\,,
$$
where the viscous stress tensor $\Pi^{\mu \nu}$ contains correction terms to the ideal
energy-momentum tensor due to shear and bulk viscosity. 
At this point, I recall that for vanishing charge density\footnote{ 
For non-vanishing charge density $\rho$, the presence of the charge current $j^\mu$ 
offers other choices to define $u^\mu$, such as $u_\mu j^\mu = \rho$. In the Landau-Lifshitz
frame, since $u^\mu$ is defined via the rest-frame of the energy-density, 
heat diffusion does not exist; but, since in this frame generally $u_\mu j^\mu \neq \rho$, there
can be charge diffusion. In the so-called Eckart frame $u_\mu j^\mu = \rho$, and 
there is heat diffusion instead of charge diffusion, indicating that these concepts are different
manifestations of the same phenomenon.}
the only useful way
to define the fluid velocity $u^\mu$ is the Landau-Lifshitz condition
$$
u_\mu T^{\mu \nu}=\varepsilon\  u^\nu\,,
$$
which implies $u_\mu \Pi^{\mu \nu}=0$.
The viscous stress tensor is customarily separated into 
a traceless part ($\pi^{\mu\nu}$) and a part with non-vanishing trace ($\Pi$),
$$
\Pi^{\mu \nu}=\pi^{\mu \nu}+\Delta^{\mu \nu} \Pi\,.
$$

For conformal fluids, where $T^{\mu}_\mu=0$, the trace part vanishes
identically and the structure of the traceless part $\pi^{\mu \nu}$ 
in the fluid  dynamic (small gradient) regime is generated by $\sigma^{\mu\nu}$ 
and the tensors found in Eq.~(\ref{conftens}) (cf. Ref.~\cite{Baier:2007ix}).
For general fluids, also the tensors in Eq.~(\ref{nctens}) contribute.
Using instead of ${\cal O}_1^{\mu \nu}$ the expression\footnote{
Note the wrong sign of the ${\cal O}_5$ term in Ref.~\cite{Baier:2007ix} that has been corrected here.}
\beq
\left<D\sigma^{\mu \nu}\right> +\frac{2-3c_s^2}{3} \sigma^{\mu \nu}\mt 
\simeq{\cal O}_1^{\mu \nu}-{\cal O}_2^{\mu \nu}-\frac{1}{2}{\cal O}_3^{\mu \nu}
+2 {\cal O}_5^{\mu \nu}-2 \frac{d\, c_s^2}{d \ln \sn} \npd^{<\mu} \ln \sn \npd^{\nu>} \ln\sn\,,
\label{dsigmamunu}
\eeq
which is accurate to second order in gradients and amounts to 
a particular resummation of higher order terms, one finds
\bqa
\pi^{\mu \nu}&=&-\eta \sigma^{\mu \nu}+\eta\tau_\pi \left[\left< D \sigma^{\mu \nu}\right>+\frac{\nabla\cdot u}{3} \sigma^{\mu \nu}\right]
+\kappa\left[R^{<\mu \nu>}-2 u_\alpha u_\beta R^{\alpha <\mu \nu>\beta}\right]
\nonumber\\
&&+\lambda_1 \sigma^{<\mu}_{\quad \lambda}\sigma^{\nu> \lambda}+\lambda_2 \sigma^{<\mu}_{\quad \lambda} \Omega^{\nu> \lambda}
+\lambda_3 \Omega^{<\mu}_{\quad \lambda}\Omega^{\nu> \lambda}\nonumber\\
&&+\kappa^* 2 u_\alpha u_\beta R^{\alpha <\mu \nu> \beta} + \eta \tau_\pi^* \frac{\nabla\cdot u}{3} \sigma^{\mu \nu}
+\lambda_4 \nabla^{<\mu} \ln \sn \nabla^{\nu>}\ln \sn\,.
\label{pimunugen}
\eqa

The coefficient of the first order gradient term is the familiar shear viscosity
coefficient $\eta$. The coefficients 
$\tau_\pi,\tau_\pi^*,\kappa,\kappa^*,\lambda_1,\lambda_2,\lambda_3,\lambda_4$
are ``second order'' transport coefficients, three of which ($\tau_\pi^*,\kappa^*,\lambda_4$)
must be identically zero for conformal fluids since they multiply structures that do
not transform homogeneously under Weyl rescalings.

The expression for the trace part $\Pi$ in the fluid dynamic regime contains all scalars up to second order
in gradients, which are given by Eq.~(\ref{confscals}) and Eq.~(\ref{ncscals}). 
Using instead of ${\cal S}_3$ the expression
$$
D \mt \simeq -\frac{1}{4} {\cal S}_1+{\cal S}_2-{\cal S}_3-\frac{1}{6}{\cal S}_4
-\frac{1}{4}{\cal S}_5+\left(\frac{3 c_s^4}{2}-\frac{d\, c_s^2}{d\ln \sn}\right){\cal S}_6-\frac{3}{2}{\cal S}_7\,,
$$ 
one finds
\bqa
\Pi&=&-\zeta\left(\nabla \cdot u\right)+\zeta \tau_\Pi D\mt
+\xi_1 \sigma^{\mu \nu} \sigma_{\mu \nu}+\xi_2 \left(\nabla \cdot u\right)^2
\nonumber\\
&&+\xi_3 \Omega^{\mu \nu} \Omega_{\mu \nu}
+\xi_4 \npu_\mu \ln \sn\npd^\mu \ln \sn+\xi_5 R
+\xi_6 u^\alpha u^\beta R_{\alpha \beta}\,.
\label{Pigen}
\eqa
Here $\zeta$ is the familiar bulk viscosity coefficient and $\tau_\Pi,\xi_1,\xi_2,\xi_3,\xi_4,\xi_5,\xi_6$
are second order transport coefficients for non-conformal fluids. As will be shown below, 
at least two of these second order coefficients (e.g., $\xi_5,\xi_6$) are completely 
specified in terms of other
transport coefficients.
The equations (\ref{pimunugen}),(\ref{Pigen}) give the most general
structure for the energy-momentum tensor of a relativistic viscous 
fluid at zero charge density up to second order in gradients.

Similar to the case for ideal fluids, the equations of motion for a viscous 
fluid are given by $\nabla_\mu T^{\mu\nu}=0$. In particular, using again 
the basic thermodynamic relations one finds for $u_\nu \nabla_\mu T^{\mu\nu}=0$ the result
\beq
Ds+s\mt= - \frac{1}{2 T}\pi^{\mu\nu}\sigma_{\mu \nu}-\frac{1}{T}\Pi \mt\,,
\label{visceoms}
\eeq
which will be useful in the following.

\subsection{Dispersion Relations and Kubo Formula}

Considering small perturbations $\delta \varepsilon, \delta u^\mu$ 
around 
an equilibrium configuration in flat space with $\varepsilon={\rm const.}$  
and $u^\mu=\left(1,{\bf 0}\right)$, 
one obtains dispersion
relations for the collective modes along ($L$) and perpendicular ($T$) to the 
perturbation.
The analysis is straightforward (see for example Ref.~\cite{Romatschke:2009im} \S II.C)
and in the fluid dynamic regime ($\omega,k\ll 1$) leads to
\bqa
\omega_{L}(k)&=&\pm k c_s -i k^2 \Gamma
\mp\frac{k^3}{2 c_s} \left[
\Gamma^2
-2 c_s^2 \left( \frac{2}{3} \frac{\eta\, \tau_\pi}{\varepsilon+P} +
\frac{1}{2}\frac{\zeta\,\tau_\Pi }{\varepsilon+P} \right)\right]
+{\cal O}(k^4)\,,\nonumber\\
\omega_{T}(k)&=&-\frac{i \eta k^2}{\varepsilon+P}+{\cal O}(k^4)\,,
\eqa
where it is recalled that transverse and longitudinal perturbations
correspond to the shear and sound mode, respectively, and
the ``sound attenuation length'' is given by
$$
\Gamma=\left(\frac{2}{3} \frac{\eta}{\varepsilon+P}
+\frac{1}{2}\frac{\zeta}{\varepsilon+P}\right)\,.
$$

The equations of motion for a viscous fluid are found to be causal 
if the maximal propagation speeds at high wavenumber $k$ for the sound 
and shear mode are smaller than the speed of light
(cf.~\cite{Romatschke:2009im} \S II). Using the dispersion 
relations, one finds for the propagation speeds in natural units
($\hbar=c=k_B=1$):
\beq
\lim_{k\rightarrow \infty} \frac{d\,\omega_T(k)}{d\,k}=
\sqrt{\frac{\eta}{\tau_\pi(\varepsilon+P)}}\,,\quad
\lim_{k\rightarrow \infty} \frac{d\,\omega_L(k)}{d\,k}=\sqrt{c_s^2+\frac{4}{3}\frac{\eta}{\tau_\pi(\varepsilon+P)}+\frac{\zeta}{\tau_\Pi(\varepsilon+P)}}\,.
\label{vTvL}
\eeq

The values are fixed by the first and second order transport coefficients,
or the properties of system in the fluid dynamic (long wavelength) limit, while
causality concerns the property of the system for short wavelength perturbations.
There is a priori no reason why $\lim_{k\rightarrow \infty} \frac{d\,\omega_{T,L}}{d\,k}$ should be less than unity\footnote{
One could still use the second order equations of motion as a phenomenological model
of a relativistic viscous fluid if the above conditions were violated for a particular
quantum field theory: adjusting the second order transport coefficients $\tau_\pi,\tau_\Pi$
``by hand'' to repair causality,  the resulting model would still treat the first order gradients correctly.
In a way, something similar is done for numerical simulations of ideal fluids: to 
dampen the turbulent instability inherent to ideal fluid dynamics, one has to introduce
``numerical viscosity'' (first order gradient correction terms) to obtain stable evolution 
for the ideal (zeroth order gradient) fluid. 
Here second order gradients are needed to obtain stable and causal evolution
of a viscous (first order gradient) fluid.}.

The second order transport coefficients are related
to thermal correlators via general Kubo formulas. As an example,
let us calculate the retarded correlator $G^{xy,xy}_R$ for 
the energy momentum tensor component $T^{xy}$. Considering 
a metric perturbation $\delta g_{\mu\nu}$
with only the non-vanishing component $\delta g_{xy}(t,z)$,
the fluid stays at rest and in equilibrium: 
$\epsilon={\rm const},u^\mu=\left(1,{\bf 0}\right)$,
since this corresponds to a tensor perturbation.
The correlator may then be found from the linear response
of $T^{xy}$ to this perturbation (cf.~\cite{Baier:2007ix}):
$$
G^{xy,xy}_R=P-i \eta\, \omega+\left(\eta\, \tau_\pi
-\frac{\kappa}{2}+\kappa^*\right) \omega^2
-\frac{\kappa}{2}k^2 + {\cal O}(\omega^3,k^3)\,.
$$
This implies that once $\tau_\pi$ is known, both $\kappa$ and $\kappa^*$
can be found by calculating this correlator for a quantum
field theory in the appropriate limit.

\section{Non-Equilibrium Entropy}
\label{sec:neentropy}

For a system in equilibrium the entropy current is simply given by the product of
entropy density and fluid velocity, $S^\mu_{\rm eq}=s u^\mu$. For systems
out of equilibrium it may be that the entropy current gets modified\footnote{I am not
aware of any experimental verification of this hypothesis.}. It is known
from kinetic theory that at zero chemical potential 
(in the absence of heat/charge diffusion),
the first correction to the equilibrium entropy current must be of second order in
gradients \cite{RKT}. In the past, the form of the non-equilibrium entropy current
has often be postulated \cite{IS0}. More recently, a more fundamental approach
has been advocated \cite{Loganayagam:2008is,Bhattacharyya:2008xc} 
that calls for all structures in a gradient expansion to be allowed.
I will follow this approach here, recovering and extending  
some of the results in Ref.~\cite{Bhattacharyya:2008xc}.

\subsection{Conformal Fluids}
\label{ssconf}

For conformal fluids, the entropy current must be built out of elements
that are invariant under conformal transformations, which are
the three scalars (\ref{confscals}) and two vectors (\ref{confvec}):
\bqa
S^{\mu}_{\rm non-eq}&=&s u^\mu+\frac{A_1}{4} {\cal S}_1 u^\mu + A_2 {\cal S}_2 u^\mu + 
A_3\left(4 {\cal S}_3 - \frac{1}{2}{\cal S}_1 +2 {\cal S}_2\right) u^\mu\nonumber\\
&&\hspace*{2.5cm}+ B_1 \left(\frac{1}{2}{\cal V}_1^\mu +\frac{u^\mu}{4} {\cal S}_1\right)+ B_2 
\left({\cal V}_2^\mu-u^\mu {\cal S}_2\right)\,,
\label{ecc}
\eqa
where the five coefficients $A_{1,2,3}$ and $B_{1,2}$ are (mass dimension one) functions of 
entropy only and the combinations and prefactors have been chosen such as to facilitate comparison to
 Ref.~\cite{Bhattacharyya:2008xc}. For conformal fluids in three dimensions one has
$c_s^2=\frac{1}{3}$ and $\Pi=0$, and the absence of a second dimensionful scale 
leads to the relation $A_{i},B_{i}\sim s^{1/3}$. 
According to Boltzmann's H-theorem, entropy is never allowed
to decrease, so the divergence of the non-equilibrium entropy current should obey 
$$\nabla_\mu S^{\mu}_{\rm non-eq}\ge 0\,.$$

The divergence of the entropy current is a physical observable, and
as such should transform homogeneously under Weyl rescalings.
Explicitly, one can convince oneself that this is the case by
writing
$$
\nabla_\mu S^{\mu}=\frac{1}{\sqrt{-g}}\partial_\mu \left(\sqrt{-g} S^\mu\right)
\rightarrow \frac{e^{4 w}}{\sqrt{-g}} 
\partial_\mu \left(e^{-4 w} \sqrt{-g}\ e^{4 w} S^\mu\right)\,.
$$

Taking the covariant derivative of Eq.~(\ref{ecc}), the result for the equilibrium 
part $\nabla_\mu \left(s u^\mu\right)$ can be read off from Eq.~(\ref{visceoms}).
A somewhat more lengthy calculation (see appendix \ref{sec:appA} for some useful identities) gives 
\bqa
\label{entropyc}
\nabla_\mu S^\mu_{\rm non-eq}&=&\frac{1}{2}\npu_{\mu}\npu_\nu \sigma^{\mu \nu}
\left(-2 A_3+ B_1\right)
+\frac{1}{3}\npu_\mu \sigma^{\mu\nu}\npu_\nu \ln \sn \left(-2A_3+ B_1\right)\nonumber\\
&&\hspace*{-2cm}+\sigma_{\mu \nu} \left[\frac{\eta}{2T} \sigma^{\mu \nu}+R^{\mu \nu} \left(
-\frac{\kappa}{2 T}+A_3\right)
+u_\alpha u_\beta R^{\alpha <\mu \nu> \beta}\left(\frac{\kappa-\eta\tau_\pi}{T}
+A_1+B_1-2A_3\right)\right.\nonumber\\
&&\hspace*{-2cm}\left.-\frac{1}{4}\sigma^{\mu}_{\ \lambda} \sigma^{\nu \lambda}\left(
\frac{2\lambda_1-\eta \tau_\pi}{T}
+A_1+B_1-2A_3\right)
+\frac{1}{3}\npd^{<\mu} \npd^{\nu>}\ln \sn\left(\frac{\eta \tau_\pi}{T}-A_1-2A_3\right)
\right.\nonumber\\
&&\hspace*{-2cm}\left.+\Omega^\mu_{\ \alpha} \Omega^{\nu \alpha}
\left(-\frac{\lambda_3+2 \eta \tau_\pi}{2T}+A_1-2A_2-2A_3+B_1\right)
\right.\\ 
&&\hspace*{-2cm}\left.
+\sigma^{\mu \nu}\frac{\mt}{12}\left(\frac{2 \eta \tau_\pi}{T}-2A_1+6 A_3-5 B_1\right)
+\frac{1}{9}\npd^{<\mu} \ln \sn \npd^{\nu>}\ln \sn\left(-\frac{\eta \tau_\pi}{T}+A_1+B_1\right)
\right]\,,\nonumber
\eqa
where the conformal fluid property $\tau_\pi^*=\kappa^*=\lambda_4=0$ was used.
Note that $B_2$ completely drops out the the divergence of the entropy current because
the covariant derivative of the relevant term is vanishing.
Positivity of $\nabla_\mu S^{\mu}_{\rm non-eq}$ is usually guaranteed by the term 
$\sigma_{\mu\nu}\sigma^{\mu \nu}$, which is of second order in gradients and hence
generally much larger than all the other (third order gradient) terms. However,
there is the possibility that $\sigma_{\mu \nu}$ itself is accidentally small,
e.g., when considering a fluid velocity field that has very little shear motion in it.
In this case it can happen that third order gradients dominate the entropy production,
and hence their coefficients must be such that entropy never decreases.
This immediately implies
\beq
B_1=2 A_3\,,
\label{B1cond}
\eeq
since the first two terms of Eq.~(\ref{entropyc}) could otherwise lead to negative
entropy production (this relation was already pointed out in Ref.~\cite{Bhattacharyya:2008xc}).
Now by the same logic, one would also expect the coefficients of the 
term $\sigma_{\mu \nu}R^{\mu \nu}$ 
to vanish, leading to
\beq
A_3=\frac{\kappa}{2T}\,.
\label{A3cond}
\eeq
However, there is a possible loophole in the argument leading to Eq.(\ref{A3cond}): 
it could be that the term $\sigma_{\mu \nu}R^{\mu \nu}$ 
combines with other terms of second and fourth order in gradients
to form a full square, e.g.,
\bqa
\sigma_{\mu \nu} \left[\frac{\eta}{2 T} \sigma^{\mu \nu}+ R^{\mu \nu} \left(A_3-\frac{\kappa}{2T}\right)\right]
&=&-\frac{T}{2 \eta} \left(A_3-\frac{\kappa}{2T}\right)^2 R_{\mu \nu} R^{\mu \nu}
\nonumber\\
&&\hspace*{-4cm}
+\frac{\eta}{2T} \left[\sigma_{\mu \nu}+\frac{T}{\eta}\left(A_3-\frac{\kappa}{2T}\right) R_{\mu \nu}\right]
\left[\sigma^{\mu \nu}+\frac{T}{\eta}\left(A_3-\frac{\kappa}{2T}\right) R^{\mu \nu}\right]
\,.
\label{scenario}
\eqa
In this scenario (pointed out in Ref.~\cite{Bhattacharyya:2008xc}), 
the coefficient $A_3$ could be arbitrary and positivity of the entropy current
would still be guaranteed if the offending first term on the r.h.s of Eq.~(\ref{scenario}) is offset by 
another term of fourth order in gradients. There are three possible sources of fourth order gradient terms in 
Eq.~(\ref{entropyc}): third order gradients in $\pi^{\mu \nu}$ times $\sigma_{\mu \nu}$ 
stemming from $\nabla_\mu \left(s u^\mu\right)$, first order gradients in $\pi^{\mu\nu}$
times third order gradients stemming from viscous corrections to Eq.~(\ref{eqds}),
and finally derivatives of third order gradients in the non-equilibrium entropy current $S^\mu_{\rm non-eq}$.
First note that the term $R_{\mu \nu}R^{\mu \nu}$ cannot be offset by a fourth
order gradient term from $\nabla_\mu \left(s u^\mu\right)$ because 
--- as Eq.~(\ref{visceoms}) shows --- the latter always
involves $\sigma_{\mu \nu}$ which is not expressible in terms of $R_{\mu \nu}$.
Secondly, viscous corrections to Eq.~(\ref{eqds}) up to second
order in gradients are found to be
\bqa
D \ln \sn &\simeq&-\mt +\eta \frac{\sigma^{\mu \nu} \sigma_{\mu \nu}}{2 s T}\,,\nonumber\\
D u^\alpha &\simeq&-c_s^2 \npd^\alpha \ln \sn
+\frac{\npu_\mu \left(\eta \sigma^{\mu \alpha}\right)}{s T}
-c_s^2 \npu_\mu \ln \sn \frac{\eta \sigma^{\mu \alpha}}{s T}-
u^\alpha \eta
\frac{\sigma^{\mu \nu}\sigma_{\mu \nu}}{2 s T}\,,
\eqa
which shows that ---again--- the term $R_{\mu \nu}R^{\mu \nu}$ cannot be offset by
these contributions. Finally, all third order gradient terms contributing
to the entropy current may be found by the same principle as in section \ref{sec:setup}.
Explicitly, one finds as scalars: 
\begin{itemize}
\item 
$\sigma_{\mu \nu}$ times all symmetric second order tensors 
(e.g., $\sigma_{\mu \nu}\,{\cal O}_{1}^{\mu \nu}$), 
\item 
$\npu_\mu \ln \sn$ times all second order vectors (e.g., $\npu_\mu \ln \sn\,{\cal V}_{1}^{\mu}$), 
\item 
$\mt$ times all second order scalars (e.g., $\mt\, {\cal S}_{1}$),
\item
$\npd^\mu$ acting on all second order vectors (e.g., $\npu_\mu {\cal V}_1^\mu$),
\end{itemize}
and as vectors: 
\begin{itemize}
\item
$\npu_\mu \ln \sn$ times all second order tensors (e.g., $\npu_\alpha \ln \sn\, {\cal O}_1^{\mu \alpha}$), 
\item
$\mt,\sigma_{\mu \nu},\Omega_{\mu\nu}$ times all second order vectors 
(e.g., $\mt\, {\cal V}_1^\mu$, $\sigma_{\ \alpha}^{\mu}\, {\cal V}_1^\alpha$),
\item
$\npu_\mu \ln \sn$ times all second order scalars (e.g., $\npd^\mu \ln \sn\, {\cal S}_1$),
\item
$\npd^\mu$ acting on all second order scalars (e.g., $\npd^\mu {\cal S}_1$),
\item
$\npd^\mu$ acting on all second order tensors (e.g., $\npu_\alpha {\cal O}_1^{\alpha \mu}$)\,.
\end{itemize}
One can readily convince oneself that derivatives of third order vectors
in the entropy current will not be able to generate a term such as $R_{\mu \nu} R^{\mu\nu}$.
For scalars, the only potentially dangerous term is the time derivative of 
$\sigma_{\mu \nu}\, {\cal O}^{\mu \nu}_2$,
because ${\cal O}^{\mu \nu}_2$ contains $R_{\mu \nu}$ and $D \sigma_{\mu \nu}$ could
potentially contain another $R_{\mu \nu}$. But the explicit expression (\ref{dsigmamunu})
for $D \sigma_{\mu \nu}$ shows that this is not the case, hence no $R_{\mu \nu} R^{\mu \nu}$
term in $\nabla_\mu S^\mu_{\rm non-eq}$ is generated from any of the possible sources.

As a consequence, the scenario (\ref{scenario}) cannot be possibly realized for every 
$A_3$. Only the condition (\ref{A3cond}) ensures positivity of the non-equilibrium
entropy. 

A similar (but slightly more complicated) argument can be made for 
the terms $u_\alpha u_\beta R^{\alpha <\mu \nu> \beta} u^\gamma u^\delta R_{\gamma <\mu \nu> \delta}$
and $\Omega^{<\mu \alpha} \Omega^{\nu>}_{\ \ \alpha} \Omega_{<\mu \beta} \Omega_{\nu>}^{\ \ \beta}$:
for these terms to be canceled or made positive definite, one has to add the two third-order
gradient terms $x_1 \sigma_{\mu \nu}\, \left({\cal O}_{1}^{\mu \nu}-{\cal O}_2^{\mu \nu}\right)$
and $x_2 \sigma_{\mu \nu}\, 2 {\cal O}_{5}^{\mu \nu}$ to $S^\mu_{\rm non-eq}$ with the requirements
$$
x_1\ge \frac{T}{8 \eta} \left(\frac{\kappa-\eta \tau_\pi}{T}+A_1\right)^2\,,\qquad
x_2\ge \frac{T}{8 \eta} \left(-\frac{\lambda_3+2\eta \tau_\pi}{2T}+A_1-2 A_2\right)^2\,.
$$
To cancel the cross-term $u_\alpha u_\beta R^{\alpha <\mu \nu> \beta} \Omega^{<\mu \gamma}\Omega_{\nu>}^{\ \ \gamma}$,
however, implies the condition
$$
x_1+x_2=\frac{T}{4\eta}  \left(\frac{\kappa-\eta \tau_\pi}{T}+A_1\right)
\left(-\frac{\lambda_3+2\eta \tau_\pi}{2T}+A_1-2 A_2\right)\,.
$$
The resulting inequality relation can only be fulfilled if 
\beq
A_2=-\frac{2 \kappa+\lambda_3}{4 T}\,.
\label{A2cond}
\eeq

Unfortunately, I did not find an argument that would fix the value of $A_1$
in terms of second order transport coefficients. Nevertheless, it seems that
a bound on the value of $A_1$ could be found by extending 
$\pi^{\mu \nu}$ to third order in gradients, 
since $x_1 \sigma_{\mu \nu} D \left({\cal O}_1^{\mu \nu}-{\cal O}_2^{\mu \nu}\right)$
involves $D \left(u_\alpha u_\beta R^{\alpha <\mu \nu> \beta}\right)$ which
seemingly has to be canceled exactly by a corresponding expression in $\pi^{\mu \nu}$,
fixing $x_1$. 

To conclude, requiring positivity of the divergence of the non-equilibrium entropy current
fixes three of the five possible coefficients ($A_2,A_3,B_1$). The divergence then takes the form
\bqa
\nabla_\mu S^\mu_{\rm non-eq}&=& \frac{\eta}{2T}\sigma_{\mu \nu}  \sigma^{\mu \nu}
+ \frac{\kappa-2 \lambda_1}{4 T}\sigma_{\mu \nu} \sigma^{\mu}_{\ \lambda} \sigma^{\nu \lambda}
\nonumber\\
&&+\left(\frac{A_1}{2}+\frac{\kappa-\eta \tau_\pi}{2 T}\right) \sigma_{\mu \nu}
\left[\left<D \sigma^{\mu \nu}\right>+\frac{1}{3}\sigma^{\mu \nu}\mt\right]\,,
\label{divec}
\eqa
which fixes the rate of entropy production slightly out of equilibrium for any
conformal fluid up to one unknown parameter ($A_1$). Note that for any theory
with $\kappa\neq 2\lambda_1$, the divergence of the non-equilibrium entropy current 
must differ from the simple expectation $\nabla_\mu S^\mu=\frac{\eta}{2T}\sigma_{\mu \nu} \sigma^{\mu \nu}$.

\subsection{Non-Conformal Fluids}
\label{sec:ncfen}

For non-conformal fluids, the non-equilibrium entropy current may be built out of
all gradient structures. Keeping the form of the conformal entropy current intact,
up to second order in gradients this leads to
\bqa
S^{\mu}_{\rm non-eq}&=&s u^\mu+\frac{A_1}{4} {\cal S}_1 u^\mu + A_2 {\cal S}_2 u^\mu + 
A_3\left(4 {\cal S}_3 - \frac{1}{2}{\cal S}_1 +2 {\cal S}_2\right) u^\mu
+A_4 {\cal S}_4u^\mu+A_5 {\cal S}_5u^\mu\nonumber\\
&&\hspace*{-1cm}+A_6 {\cal S}_6 u^\mu+ B_1 \left(\frac{1}{2}{\cal V}_1^\mu +\frac{u^\mu}{4} {\cal S}_1\right)+ B_2 
\left({\cal V}_2^\mu-u^\mu {\cal S}_2\right)
+B_3 \left({\cal V}_3^\mu-\frac{1}{2}{\cal V}_1^\mu\right)+B_4 {\cal V}_4^\mu\,
\nonumber\\
&&\hspace*{-1cm}+B_5 {\cal V}_5^\mu+B_6 {\cal V}_6^\mu+\frac{A_7}{4}\left(\frac{2}{c_s^2} {\cal S}_7 u^\mu +\frac{4}{c_s^2}{\cal S}_3u^\mu -\frac{2}{3 c_s^2}{\cal S}_4u^\mu 
+\frac{1}{c_s^2}{\cal S}_5u^\mu  +2 c_s^2{\cal S}_6u^\mu
+3{\cal V}_1^\mu\right.\nonumber\\
&&\hspace*{4cm}+6{\cal V}^\mu_2-6 {\cal V}_3^\mu+\left.(4-6c_s^2) {\cal V}_4^\mu-12 c_s^2{\cal V}_5^\mu-\frac{4}{3}{\cal V}_6^\mu\right)\,,
\label{ecnc}
\eqa
where the cumbersome expression multiplying $A_7$ leads to a particularly simple divergence
(see appendix \ref{sec:appA}).

Since for non-conformal fluids $c_s^2$ may be a function of the entropy density, its derivative
$c_s^\prime\equiv \frac{d \ln c_s^2}{d \ln s}$ is in general non-vanishing. Similarly, 
the coefficient functions $A_{i},B_{i}$ now may contain logarithms or any power
of the entropy density, so that
in general $A_{i}^\prime\equiv \frac{d \ln A_{i}}{d \ln s}\neq \frac{1}{3}$ (and the same for $B_{i}$).
Using Eq.~(\ref{visceoms}), the divergence of the non-equilibrium entropy current can be calculated,
but the resulting expression (\ref{entropync}) 
is suitably lengthy to be unenlightening except maybe for the expert reader, so it has been
relegated to the appendix. Nevertheless, the principle of positivity of the entropy current divergence
singles out conditions for the $A_{i},B_{i}$ as was the case for conformal fluids. One immediately finds
\bqa
&A_5=0\,,\quad\!\!
B_1=2 A_3\,,\quad\!\!
B_3=2(1-3c_s^2)A_3\,,\quad\!\!
B_4=-A_3(A_3^\prime-3c_s^2(1-2c_s^2))\,,&\\
&
B_5=B_2(B_2^\prime-c_s^2)+(B_3^\prime+c_s^2)B_3\,,\quad
B_6=2A_6-A_7(A_7^\prime-c_s^2)+\frac{4}{3}(1-3 c_s^2)A_3 B_3^\prime\,.
\nonumber
\eqa
In addition, one can again go through 
the arguments as in section \ref{ssconf} to show that $\sigma_{\mu \nu} R^{\mu\nu}$ and
$\mt R$ cannot form full squares because no other contribution to the divergence of
the entropy current could offset their negative contribution to fourth order in derivatives.
This leads to the conditions
$$
A_3=\frac{\kappa}{2T}\,,\quad
\xi_5=\frac{A_3 T}{3}(3 A_3^\prime -1) \,.
$$
Finally, requiring that the cross-terms 
$u_\alpha u_\beta R^{\alpha <\mu \nu> \beta} \Omega^{<\mu \gamma}\Omega_{\nu>}^{\,\ \gamma}$ and
$u_\alpha u_\beta R^{\alpha \beta} \Omega_{\mu \nu} \Omega^{\mu \nu}$ cancel leads to 
$$
A_2=-\frac{6 \kappa(1-2 c_s^2)-2\kappa^*+\lambda_3}{4 T}\,,\quad
\frac{\xi_6+\xi_3}{T}=-\frac{A_2}{3}(1-6c_s^2+3A_2^\prime)+2(1-3 c_s^2)A_3(2c_s^2-B_3^\prime)\,.
$$
Contrary to the situation for conformal fluids, fixing 8 of the 13 coefficient functions in
the entropy current still does not seem to lead to a simple form for $\nabla_\mu S^\mu_{\rm non-eq}$.
It may be possible that a more detailed analysis of the effect of third order gradients 
$S^\mu$, $\pi^{\mu\nu}$ and $\Pi$ could result in further conditions on either the remaining five
coefficient functions or some of the second order transport coefficients 
$\tau_\pi^*,\kappa^*,\lambda_4,\xi_1,\xi_2,\xi_3,\xi_4$.

\section{Second Order Transport Coefficients: Known Results}
\label{sec:soknown}

The most general structure for the viscous energy-momentum tensor up to second order
in gradients has been established in section \ref{sec:VEMT}. For conformal
fluids at vanishing charge density, the energy-momentum tensor --- and hence
the equations of motion for viscous fluid dynamics --- depend on 7 dimensionless
numbers: the speed of sound and the transport coefficients. To first order in gradients,
the only transport coefficient is the shear viscosity $\eta$, and to second order there
are in general five additional coefficients: $\tau_\pi,\kappa,\lambda_1,\lambda_2,\lambda_3$.
All of these transport coefficients seem to be independent, but so far $\lambda_3=0$
has been found for all examples of quantum field theories where it has been calculated,
perhaps suggesting that there is another symmetry that has not been exploited yet.

For non-conformal fluids, the energy-momentum tensor can no longer be expressed
in terms of dimensionless numbers because of the presence of an additional scale.
Combinations involving transport coefficients that have vanishing mass dimension
then are functions of, e.g., the entropy density. Also, there are 9 new 
independent transport coefficients in addition to the 6 for conformal fluids: 
to first order in gradients the bulk viscosity $\zeta$
and to second order $\tau_\Pi,\tau_\pi^*,\kappa^*,\lambda_4,\xi_1,\xi_2,\xi_3,\xi_4$,
while $\xi_5,\xi_6$ are specified in terms of the other transport coefficients and their 
derivatives.

Many studies have been published on the value of the first order transport coefficients
$\eta,\zeta$, and hence I will focus here exclusively on second order transport coefficients
and review the existing results for relativistic field theories. Of particular
interest are the values of the ``relaxation times'' $\tau_\pi,\tau_\Pi$ 
in the shear and bulk channel, respectively, since these have implications 
for numerical solutions of relativistic viscous fluid dynamics.

For conformal fluids in 3+1 spacetime dimensions, $\tau_\pi$ is known for quantum field theories at
weak coupling $\lambda\ll 1$ \cite{York:2008rr}, and for a particular quantum field example (${\cal N}=4$ SYM theory) 
at very large coupling $\lambda\gg1$ \cite{Buchel:2008bz}, 
and for infinite coupling $\lambda\rightarrow \infty$ \cite{Baier:2007ix,Bhattacharyya:2008jc}:
$$
\lim_{\lambda\rightarrow 0}\tau_\pi\sim 5.9 \frac{\eta}{\varepsilon+P}\,,\quad
\lim_{\lambda\rightarrow \infty}\tau_\pi\sim\left(4-2\ln 2+\frac{375}{8} \zeta(3) \lambda^{-3/2}
\right)\frac{\eta}{\varepsilon+P}\,.
$$
For non-conformal fluids, $\tau_\Pi$ is known only in one particular example of a strongly coupled
field theory \cite{Kanitscheider:2009as} where
$$
\tau_\Pi=\tau_\pi\,,\quad \zeta=2\eta \left(\frac{1}{3}-c_s^2\right)\,,\quad \tau_\pi=\left(4-2\ln 2\right) \frac{\eta}{\varepsilon+P}\,.
$$

Remarkably, for these cases, the maximal propagation speeds (\ref{vTvL}) 
turn out to be less than unity, indicating that 
the second order fluid dynamic equations of motion obey causality \cite{Romatschke:2009im}. 
Furthermore, it was shown recently 
that causality in second order fluid dynamics is guaranteed by
the causality of the underlying quantum field theory
for the conformal field theory  example of ${\cal N}=2$ SYM at strong coupling 
\cite{Buchel:2009tt}.
This seems to suggest that second order fluid dynamics may be a 
particularly useful approximation of quantum field theories in the appropriate limit.

The parameter $\lambda_1$ is known for quantum field theories at
weak coupling \cite{York:2008rr}, and for ${\cal N}=4$ SYM theory at 
very large, and infinite coupling \cite{Buchel:2008kd,Baier:2007ix,Bhattacharyya:2008jc}:
$$
\lim_{\lambda\rightarrow 0}\lambda_1\sim 5.2 \frac{\eta^2}{\varepsilon+P}\,,\quad
\lim_{\lambda\rightarrow \infty}\lambda_1\sim\left(2+\frac{215}{4} \zeta(3) \lambda^{-3/2}\right)\frac{\eta^2}{\varepsilon+P}\,.
$$
The parameter $\lambda_2$ is known for quantum field theories at
weak coupling \cite{York:2008rr}, and for ${\cal N}=4$ SYM theory at infinite coupling \cite{Bhattacharyya:2008jc}:
$$
\lim_{\lambda\rightarrow 0}\lambda_2\sim -2 \eta \tau_\pi\,,\quad
\lim_{\lambda\rightarrow \infty}\lambda_2\sim - \ln 2 \frac{\eta}{\pi T}\,.
$$
In attempts to calculate $\lambda_3$, the parameter has been found to vanish at weak and strong coupling 
\cite{York:2008rr,Bhattacharyya:2008jc}.

The parameter $\kappa$ has been calculated in weakly coupled $SU(N)$ gauge theory \cite{Romatschke:2009ng} and
for ${\cal N}=4$ SYM theory for large, and for infinite coupling \cite{Baier:2007ix,Buchel:2008bz}:
$$
\lim_{\lambda\rightarrow 0}\kappa\sim \frac{5 s}{8 \pi^2 T}\,,\quad
\lim_{\lambda\rightarrow \infty}\kappa\sim \frac{s}{4\pi^2 T}\left(1-\frac{145}{8} \zeta(3) \lambda^{-3/2}\right) \,.
$$
Note that $\kappa\neq 2\lambda_1$, except for ${\cal N}=4$ SYM theory at infinite coupling.

For the strongly coupled theory example of Ref.~\cite{Kanitscheider:2009as}, all second order transport coefficients for a 
particular class of non-conformal fluids may be extracted:
\bqa
&\kappa^*=-\frac{\kappa}{2 c_s^2}(1-3 c_s^2)\,,\quad
\tau_\pi^*=-\tau_\pi (1-3c_s^2)\,,\quad
\lambda_4=0\,,&\nonumber\\
&
\xi_1=\frac{\lambda_1}{3}\left(1-3c_s^2\right)\,,\quad
\xi_2=\frac{2 \eta \tau_\pi c_s^2}{3}\left(1-3c_s^2\right)\,,\quad
\xi_3=\frac{\lambda_3}{3}\left(1-3c_s^2\right)\,,\quad
\xi_4=0\,,\quad\nonumber\\
&\xi_5=\frac{\kappa}{3}\left(1-3c_s^2\right)\,,\quad
\xi_6=\frac{\kappa}{3 c_s^2}\left(1-3c_s^2\right)\,.&
\eqa
As a non-trivial consistency check, one can proceed to evaluate the relations between $\xi_5,\xi_6$ and the other
transport coefficient derived in section \ref{sec:ncfen}. Using $T=s^{c_s^2}$ 
and $\kappa\propto s/T$, one finds
\bqa
\xi_5=\frac{A_3 T}{3}(3 A_3^\prime -1)&=&\frac{\kappa}{3}(1-3c_s^2)\,,\nonumber\\
\xi_6+\xi_3=-\frac{T}{3}A_2(1-6c_s^2+3A_2^\prime)+2T(1-3 c_s^2)A_3(2c_s^2-B_3^\prime)&=&\frac{1-3c_s^2}{3 c_s^2}
\left(\kappa+\lambda_3 c_s^2\right)\,,\nonumber
\eqa
which precisely matches the values from Ref.~\cite{Kanitscheider:2009as}. 
It would be interesting to check these relations for other examples
of non-conformal field theories (e.g., by compactification
of conformal field theories suggested in \cite{Buchel:2007mf}).

Furthermore, one can use the above results for the conformal fluid transport coefficients for 
infinitely strongly coupled ${\cal N}=4$ SYM theory and evaluate the
three relations (\ref{B1cond}),(\ref{A3cond}),(\ref{A2cond}):
\beq
\frac{(\pi T)^2}{4 \pi \eta} A_2=-\frac{1}{8}\,,\quad
\frac{(\pi T)^2}{4 \pi \eta} A_3=\frac{1}{8}\,,\quad
\frac{(\pi T)^2}{4 \pi \eta} B_1=\frac{1}{4}\,.
\label{agree}
\eeq
These values precisely correspond to those for the entropy current
derived from the horizon of a black hole in $AdS_5$ (see Ref.~\cite{Bhattacharyya:2008xc}). 
This serves as another consistency-check of the above approach.

It would be interesting to calculate the entropy current from the black hole horizon
corresponding to the geometry considered in Ref.~\cite{Kanitscheider:2009as}
and check that its form fulfills the conditions on $B_1,B_3,B_4,B_5,B_6,A_2,A_3,A_5$
found in section \ref{sec:ncfen}.

Also, it could be possible to constrain the coefficient $A_1$ 
for ${\cal N}=4$ SYM using existing results in a gradient expansion
beyond second order \cite{Lublinsky:2009kv}.

\section{Discussion and Conclusions}
\label{sec:conclusions}

In this work, I have derived the structure of the energy-momentum tensor and non-equilibrium entropy
current for a relativistic viscous fluid in curved space at vanishing charge density up to second order in 
gradients. I found that in this case there are 15 possible transport coefficients multiplying
second order gradient terms (10 of which vanish for conformal fluids), 
in addition to the bulk and shear viscosity coefficients arising at first order.
Requiring the divergence of the (non-equilibrium) entropy current to be positive definite, two of these 15 coefficients are 
found not to be independent, and their relation to the other coefficients is specified.

Also, the most general form of the entropy current out of equilibrium allows 13 possible coefficient functions
(8 of which vanish for conformal fluids) multiplying terms of second order in gradients. 
Requiring positivity fixes 8 (conformal fluids: three) of these coefficient functions.

These results resolve any ambiguities in the structure of the viscous fluid dynamic energy-momentum tensor
and thereby clear the path for an extraction of the (second order) transport coefficients
from quantum field theories. For instance, using Eq.~(\ref{pimunugen}) and Eq.~(\ref{Pigen}) one can derive
Kubo-like formulas for correlators in the fluid dynamics regime for non-conformal fluids. In addition,
knowledge of the correlators and spectral functions to second order in gradients may
be vital to extract first order gauge theory transport coefficients from lattice Quantum-Chromodynamics 
\cite{Harvey}.

Also, knowing the structure of the viscous fluid dynamic equations may be 
required to extend conformal fluid dynamics simulations of
heavy-ion collisions \cite{Romatschke:2007mq,Dusling:2007gi,Song:2007ux,Molnar:2008xj} 
to reliably include effects from bulk viscosity.
It could furthermore help to clarify the role of non-linear viscous damping of the r-mode
instability in rotating neutron stars \cite{Andersson1998,Lindblom:1998wf,Duez:2004nf}

Moreover, the agreement (\ref{agree}) of the non-equilibrium entropy current from fluid
dynamics and that derived from the horizon area of a black hole support 
the hope of linking these two concepts in more detail, 
which might have wide ranging consequences.
In this context it is curious to note that for conformal fluids,
where one parameter ($A_1$) could not be fixed by requiring positivity of the entropy
current alone, a diffeomorphism ambiguity prevents fixing of the same parameter
on the gravity side. It is unknown whether $A_1$ (and hence the fluid 
entropy current) cannot be fixed in principle in fluid dynamics, or whether this is only a
manifestation of our lack of ingenuity (cf. the discussion in 
Ref.~\cite{Figueras:2009iu} 
for the gravitational standpoint). However, I believe $A_1$
should have a definite value, since nature must know
the amount of entropy production when taking a system out of equilibrium.

Furthermore, it seems that genuine non-equilibrium contributions to the
entropy current are necessary to ensure consistency of viscous fluid
dynamics with the requirement that entropy may never decrease. However,
it is unknown to me whether the presence of these non-equilibrium
contributions to the entropy current is an experimentally established fact.
If not, relations such as Eq.~(\ref{divec}) could potentially be
used to attempt such an experimental verification, since
it was found that the divergence of the non-equilibrium entropy current 
differs from the ``naive'' expectation
$\nabla_\mu S^\mu=\frac{\eta}{2T}\sigma_{\mu \nu} \sigma^{\mu \nu}$. 
E.g., this could be done by measuring the entropy of a conformal fluid in equilibrium, 
then subjecting the fluid to a perturbation $\sigma_{\mu \nu}$ 
(ideally one that is tuned such that 
$\left<D \sigma^{\mu \nu}\right>\simeq-\frac{1}{3}\sigma^{\mu \nu}\mt$ to eliminate
the dependence on the unknown coefficient $A_1$), 
switching off the perturbation and letting the system relax back to equilibrium,
then measuring again its entropy. Setups reminiscent of this proposal are being
pursued in cold atom experiments \cite{Luo:2007zz}, but it is unclear to me whether
these could be adapted to test for non-equilibrium entropy effects.

Finally, it should be possible to extend my analysis to include the case
of non-vanishing charge density, allowing for the presence of charge/heat 
diffusion by accounting for gradients in the chemical potential,
as well as including structures built out of the Levi-Civita symbol
that break parity. This would be the last step in determining the equations 
of motion for a general one-component relativistic viscous fluid.

\acknowledgments

I would like to thank A.~Buchel, 
S.~Minwalla, G.D.~Moore, R.~Myers, M.~Rangamani, K.~Skenderis, D.T.~Son and E.G.~Thompson 
for fruitful discussions and clarifications. This work was supported by
the US Department of Energy, grant number DE-FG02-00ER41132. 

\appendix

\section{Conformal Transformations}
\label{sec:weyl}
Here I collect the behavior of terms up to second order in gradients
under Weyl rescalings (\ref{Weylres}). For the Christoffel symbols one finds
$$
\Gamma^\lambda_{\mu \nu}\rightarrow\Gamma^\lambda_{\mu \nu}-\left(
\delta^\lambda_\nu \partial_\mu w+\delta^\lambda_\mu \partial_\nu w
-g_{\mu \nu} \partial^\lambda w\right)\,,
$$
from which the behavior of the covariant derivative of the fluid velocity
can be calculated to be
$$
\nabla_\mu u_\nu\rightarrow 
e^{-w} \left(\nabla_\mu u_\nu+u_\mu \partial_\nu w-g_{\mu \nu} D w\right)\,.
$$
This in turn implies
\bqa
\sigma_{\mu\nu}&\rightarrow& e^{-w}\sigma_{\mu\nu}\,,\nonumber\\
\Omega_{\mu\nu}&\rightarrow& e^{-w}\Omega_{\mu\nu}\,,\nonumber\\
\nabla\cdot u&\rightarrow&e^{w} \left(\nabla \cdot u-3 D w\right)\,,\nonumber
\eqa
which can be used to derive
\bqa
\npu_\nu \sigma^{\mu \nu}&\rightarrow&\npu_\nu \sigma^{\mu \nu}-2 \sigma^{\mu \nu}\npu_\nu w\,,\nonumber\\
\npu_\nu \Omega^{\mu \nu}&\rightarrow&\npu_\nu \Omega^{\mu \nu}\,.\nonumber
\eqa

Furthermore, using the Christoffel symbols one can calculate the
transformation of the Riemann tensor to be
$$
R^\lambda_{\ \mu \sigma \nu}\rightarrow
R^\lambda_{\ \mu \sigma \nu}
-\delta^\lambda_\nu \nabla_\mu \partial_\sigma w
+\delta^\lambda_\sigma \nabla_\mu \partial_\nu w
+g_{\mu \nu} \nabla^\lambda \partial_\sigma w
-g_{\mu \sigma} \nabla^\lambda \partial_\nu w\,,
$$
so that 
\bqa
u_\lambda u^\nu R^\lambda_{\ <\mu \sigma> \nu}\,&\rightarrow&
u_\lambda u^\nu R^\lambda_{\ <\mu \sigma> \nu}+ \nabla_{<\mu }\partial_{\sigma>} w\,,
\nonumber\\
R_{\mu \nu}&\rightarrow&R_{\mu \nu}+2 \nabla_{\mu} \partial_\nu w
+g_{\mu \nu} \nabla^\lambda \partial_\lambda w\,,\nonumber\\
R&\rightarrow&e^{2w} \left(R +6 \nabla^\lambda \partial_\lambda w\right)\,.
\eqa

Finally, one has
$$
\npu_\mu \ln s \rightarrow \npu_\mu \ln s+3 \npu_\mu w\,,
$$
and the above result for the Christoffel symbols implies
\bqa
\npu_{<\mu} \npu_{\nu>} \ln s &\rightarrow&\npu_{<\mu} \npu_{\nu>} \ln s+3 \npu_{<\mu} \partial_{\nu>} w
+2 \npu_{<\mu} \ln s \npu_{\nu>} w+\frac{3}{2} \sigma_{\mu \nu} D w\,,\nonumber\\
\npu_{\mu} \npd^{\mu} \ln s &\rightarrow&e^{2 w}\left(\npu_{\mu} \npd^{\mu} \ln s 
+3 \npu_\mu \partial^\mu w+3 \mt D w-\npu_\mu \ln s \npd^\mu w\right)\,.\nonumber
\eqa

\section{Useful Identities}
\label{sec:appA}

This appendix contains some useful identities
that were, e.g., used in calculating the divergence of 
the non-equilibrium entropy current.

Time derivatives of various scalars and vectors:
\bqa
\Omega_{\mu \nu} D \Omega^{\mu \nu} &=&
- \frac{2-3c_s^2}{3} \Omega_{\mu \nu} \Omega^{\mu \nu} \mt -
\Omega_{\mu \nu} \Omega^{\beta \nu} \sigma^{\mu}_{\ \beta}\,,
\nonumber\\
D \npu_\mu \ln \sn &=& - \npu_\mu \mt -\left(\frac{1}{2}\sigma_{\mu \beta}+\Omega_{\mu \beta}\right)
\npd^\beta \ln \sn+\left(c_s^2-\frac{1}{3}\right)\mt \npu_\mu \ln \sn
\nonumber\\
&&-u_\mu c_s^2 \npu_\alpha \ln \sn \npd^\alpha \ln \sn\,,
\nonumber\\
D \npu_\mu \npd^\mu \ln \sn &=&-\npu_\mu \npd^\mu \mt - u^\nu R_{\mu \nu} \npd^\mu \ln \sn\nonumber\\
&&-\npd^\mu \npd^\nu \ln \sn \left(\sigma_{\mu \nu}+\Delta_{\mu \nu} \mt \frac{2 -3c_s^2}{3} \right)
\\
&&+c_s^2 \npd^\mu \ln \sn \npd^\nu \ln \sn \left( \sigma_{\mu \nu}-\Delta_{\mu \nu}
\mt \frac{1+3 c_s^2-3\, (d\,\ln c_s^2/d\ln \sn)}{3} \right)\nonumber\\
&&+\npd^\mu \ln \sn \left(\frac{6 c_s^2-1}{3}\npu_\mu \mt-\frac{1}{2} \npd^\beta \sigma_{\beta \mu}
-\npd^\beta \Omega_{\beta \mu}\right)\,,
\nonumber\\
D \left[R+2 u^\alpha u^\beta R_{\alpha \beta}\right]&=&2 \left[\npu_\mu \left(u_\nu R^{\mu \nu}\right)
-R^{\mu \nu}\left(\frac{1}{2}\sigma_{\mu \nu}+\frac{1}{3}\Delta_{\mu \nu}\mt\right)
-2 c_s^2 u_\nu R^{\mu \nu} \npu_\mu \ln \sn\right]\,.
\nonumber
\eqa

Identities involving curvature tensors and vorticity:
\bqa
R^\lambda_{\ \mu \sigma\nu}u_\lambda&=&\left(\nabla_\nu 
\nabla_\sigma-\nabla_\sigma \nabla_\nu\right) u_\mu\,,
\nonumber\\
R_{\lambda \sigma} u^\lambda &=&2 c_s^2 \Omega_{\mu \sigma} \npd^\mu \ln \sn
+\npd^\mu \left(\npu_\sigma u_\mu\right)-D_\sigma \mt\nonumber\\
&&+c_s^2 u_\sigma \left[\npu_\mu \npd^\mu \ln \sn - \npu_\mu \ln \sn \npd^\mu \ln \sn \left(c_s^2-\frac{d\ \ln 
c_s^2}{d\ln \sn}\right)\right]\,,
\nonumber\\
\npd^\alpha u^\lambda R_{\alpha \lambda}&=&2 c_s^2 \npu_\alpha \Omega^{\mu \alpha} \npu_\mu \ln \sn
+\frac{1}{2}\npu_\alpha \npu_\mu \sigma^{\mu \alpha} - \frac{2}{3}\npu_\mu \npd^\mu \mt 
-2\Omega_{\mu \nu}\Omega^{\beta \nu} \sigma^{\mu}_{\ \beta}\nonumber\\
&&+\mt \left[\Omega_{\mu \nu}\Omega^{\mu \nu} \left(3 c_s^2-\frac{1}{3}\right)-\frac{1}{4}
\sigma_{\mu \nu}\sigma^{\mu \nu}-u_\mu u_\nu R^{\mu \nu}\right]\,,
\nonumber\\
\Omega_{\mu \nu} \npd^\mu \npd^\nu \ln s &=& - \Omega_{\mu \nu} \Omega^{\mu \nu} \mt\,,
\nonumber\\
\npu_\beta \npu_\alpha \Omega^{\alpha \beta}&=&
\frac{4-3 c_s^2}{3} \Omega_{\mu \nu} \Omega^{\mu \nu} \mt 
+2 \Omega_{\mu \nu} \Omega^{\beta \nu} \sigma^{\mu}_{\ \beta}\,.
\nonumber
\eqa

Identities for scalar ${\cal S}_7$:
\bqa
&\hspace*{-4cm}\frac{1}{4}\left(\frac{2}{c_s^2} {\cal S}_7 u^\mu +\frac{4}{c_s^2}{\cal S}_3u^\mu -\frac{2}{3 c_s^2}{\cal S}_4u^\mu 
+\frac{1}{c_s^2}{\cal S}_5u^\mu  +2 c_s^2{\cal S}_6u^\mu\right.&\\
&\left.+3{\cal V}_1^\mu+6{\cal V}^\mu_2-6 {\cal V}_3^\mu 
+(4-6c_s^2) {\cal V}_4^\mu-12 c_s^2{\cal V}_5^\mu-\frac{4}{3}{\cal V}_6^\mu\right)=&
\nonumber\\
&u^\mu \left(c_s^2 \npu_\alpha \ln \sn \npd^\alpha \ln \sn+\npu_\alpha \npd^\alpha \ln \sn\right)
+\npd^\mu \mt+\sigma^{\mu \alpha}\npu_\alpha \ln \sn -\frac{1}{3}\mt \npd^\mu \ln \sn\,.&
\nonumber
\eqa

\section{$\nabla_\mu S^{\mu}_{\rm non-eq}$ for Non-Conformal Fluids}

The explicit divergence of Eq.~(\ref{ecnc}) is
\bqa
&\nabla_\mu S^\mu_{\rm non-eq}=&\nonumber\\
&&\hspace*{-2cm}\npu_{\mu}\npu_\nu \sigma^{\mu \nu}
\frac{\left(-2 A_3+ B_1\right)}{2}
+\npu_\mu \sigma^{\mu\nu}\npu_\nu \ln \sn 
\left(-2 c_s^2 A_3+ \frac{B_1}{2}(c_s^2+B_1^\prime)-c_s^2 B_3+B_4\right)\nonumber\\
&&\hspace*{-2cm}+\frac{2}{3}\npu_\mu \npd^\mu \mt \left(2(1-3c_s^2) A_3-B_3\right)+A_5 D R\nonumber\\
&&\hspace*{-2cm}
+\npu_\alpha \Omega^{\mu \alpha} \npu_\mu \ln \sn \left(B_2 (B_2^\prime-c_s^2)+B_3 (B_3^\prime+c_s^2)-B_5\right)
\nonumber\\
&&\hspace*{-2cm}+\frac{1}{3}\npd^\mu \mt \npu_\mu \ln \sn \left(-4c_s^2 (1-3c_s^2) A_3-6 A_6+3 A_7(A_7^\prime-c_s^2)
-2B_3(B_3^\prime-c_s^2)+3B_6\right)
\nonumber\\
&&\hspace*{-2cm}+\sigma_{\mu \nu} \left[\frac{\eta}{2T} \sigma^{\mu \nu}+R^{\mu \nu} \left(
-\frac{\kappa}{2 T}+A_3\right)
+u_\alpha u_\beta R^{\alpha <\mu \nu> \beta}\left(\frac{\kappa-\kappa^*-\eta\tau_\pi}{T}
+A_1+B_1-2A_3\right)\right.\nonumber\\
&&\left.\hspace*{-2cm}+\Omega^\mu_{\ \alpha} \Omega^{\nu \alpha}
\left(-\frac{\lambda_3+2 \eta \tau_\pi}{2T}+A_1-2A_2-2A_3+B_1-2B_3\right)
\right.\nonumber\\ 
&&\hspace*{-2cm}\left.+c_s^2\npd^{<\mu} \npd^{\nu>}\ln \sn\left(\frac{\eta \tau_\pi}{T}-A_1-2A_3-B_3+c_s^{-2}B_4\right)
\right.\nonumber\\
&&\hspace*{-2cm}\left.-\frac{1}{4}\sigma^{\mu}_{\ \lambda} \sigma^{\nu \lambda}\left(
\frac{2\lambda_1-\eta \tau_\pi}{T}
+A_1+B_1-2A_3\right) 
+\sigma^{\mu \nu}\frac{\mt}{12}\left(\frac{2 \eta (\tau_\pi-\tau_\pi^*)-12 \xi_1+3 \zeta \tau_\Pi}{T}
\right.\right.\nonumber\\&&\hspace*{0cm}\left.\left.
-A_1(1+3A_1^\prime)+A_3(4+6A_3^\prime)-6A_4- B_1(4+3B_1^\prime)\right)
\right.\nonumber\\ 
&&\hspace*{-2cm}\left.
+\npd^{<\mu} \ln \sn \npd^{\nu>}\ln \sn\left(-\frac{2c_s^2(c_s^2-c_s^\prime)\eta \tau_\pi+\lambda_4}{2T}+
c_s^2(c_s^2-c_s^\prime)A_1+2 c_s^2 c_s^\prime A_3-A_6\right.\right.\nonumber\\
&&\hspace*{-2cm}\left.\left.+A_7(A_7^\prime-c_s^2)+c_s^2B_1^\prime B_1+c_s^2(c_s^2- B_3^\prime-c_s^\prime)B_3+(B_4^\prime-c_s^2)B_4\right)
\right]\,.\nonumber\\
&&\hspace*{-2cm}+\mt \left[\frac{\zeta}{T} \mt
+R \left(-\frac{\xi_5}{T}+(1-A_5^\prime)A_5+\frac{1}{3}(3A_3^\prime-1)A_3\right)\right.\nonumber\\
&&\hspace*{-2cm}\left.+ u_\alpha u_\beta R^{\alpha \beta}\left(-\frac{\xi_6-\zeta \tau_\Pi}{T}
-\frac{2}{3}(1-3 A_3^\prime)A_3-2A_4\right)\right.\nonumber\\
&&\hspace*{-2cm}\left.+\mt^2\left(\frac{\zeta \tau_\Pi-3\xi_2}{3T}+\frac{2}{9}(1-3A_3^\prime)A_3+\frac{1}{3}(1-3A_4^\prime)A_4\right)\right.\\
&&\hspace*{-2cm}\left.+\Omega_{\alpha \beta}\Omega^{\alpha \beta}\left(-\frac{\xi_3+\zeta \tau_\Pi}{T}
-\frac{1}{3}(1-6c_s^2+3 A_2^\prime)A_2+\frac{2}{3}(2-3c_s^2-3A_3^\prime)A_3+2A_4\right.\right.\nonumber\\
&&\left.\left.-(c_s^2-B_2^\prime)B_2
-\frac{1}{3}(1-9c_s^2)B_3-B_5\right)\right.\nonumber\\
&&\hspace*{-2cm}\left.+\npu_\mu \npd^\mu \ln \sn\  c_s^2\left(\frac{\zeta \tau_\Pi}{T}+4(c_s^2-c_s^\prime-A_3^\prime)A_3
-2A_4+(1-c_s^{-2}A_7^\prime)A_7+c_s^{-2} B_6\right)\right.\nonumber\\
&&\hspace*{-2cm}\left.+\npu_\mu \ln \sn \npd^\mu \ln \sn \left(-\frac{c_s^2(c_s^2-c_s^\prime)\zeta \tau_\Pi+\xi_4}{T}
+\frac{2}{3}c_s^2(c_s^2-2 c_s^\prime-3 c_s^2 A_3^\prime)A_3\right.\right.\nonumber\\
&&\hspace*{-2cm}\left.\left.
+2c_s^2(c_s^2-c_s^\prime)A_4+\frac{1}{3}(1+6 c_s^2-3A_6^\prime)A_6+\frac{1}{3}(1+3c_s^2)(c_s^2-A_7^\prime)A_7-(c_s^2-B_6^\prime)B_6\right)
\right]\nonumber\,,
\label{entropync}
\eqa
where it is recalled that $c_s^\prime=\frac{d \ln c_s^2}{d \ln \sn}$, $A_i^\prime=\frac{d \ln A_i}{d \ln \sn}$, and $B_i^\prime=\frac{d \ln B_i}{d \ln \sn}$.

\end{document}